\documentclass[11pt,prl,reprint,twocolumn,notitlepage,nofootinbib]{revtex4-1}

\usepackage{graphicx}
\usepackage{savesym}
\usepackage{listings}
\usepackage[intlimits]{amsmath}
\savesymbol{iint}
\usepackage{txfonts}
\usepackage{bm}
\restoresymbol{TXF}{iint}
\usepackage{amssymb}
\usepackage{hyperref}

\def\be{\begin{equation}}
\def\ee{\end{equation}}
\def\ba{\begin{eqnarray}}
\def\ea{\end{eqnarray}}
\def\go{\mathrel{\raise.3ex\hbox{$>$}\mkern-14mu
             \lower0.6ex\hbox{$\sim$}}}
\def\lo{\mathrel{\raise.3ex\hbox{$<$}\mkern-14mu
             \lower0.6ex\hbox{$\sim$}}}

\def\apj{Astrophys. J.}

\def\aap{Astron. Astrophys. }

\def\apss{Astrophys. Space Sci.}

\def\mnras{Mon. Not. Roy. Astron. Soc. }

\begin{document}
\title{Hall Attractor in Axially Symmetric Magnetic Fields in Neutron Star Crusts}

\author{Konstantinos N. Gourgouliatos}
\email{kostasg@physics.mcgill.ca}
\affiliation{Department of Physics, McGill University, 3600 rue University, Montreal, QC, Canada, H3A 2T8}
\author{Andrew Cumming}
\email{cumming@physics.mcgill.ca}
\affiliation{Department of Physics, McGill University, 3600 rue University, Montreal, QC, Canada, H3A 2T8}

\begin{abstract}
We have found an attractor for an axially symmetric magnetic field evolving under the Hall effect and subdominant ohmic dissipation, resolving the question of the long term fate of the magnetic field in neutron star crusts. The electron fluid is in isorotation, analogous to Ferraro's law, with its angular velocity being approximately proportional to the poloidal magnetic flux, $\Omega \propto \Psi$. This equilibrium is the long term configuration of a magnetic field evolving because of the Hall effect and ohmic dissipation. For an initial dipole dominated field the attractor consists mainly of a dipole and an octupole component accompanied by an energetically negligible quadrupole toroidal field. The field dissipates in a self-similar way: although higher multipoles should have been decaying faster, the toroidal field mediates transfer of energy into them from the lower ones, leading to an advection diffusion equilibrium and keeping the ratio of the poloidal multipoles almost constant. This has implications for the structure of the intermediate age neutron stars, suggesting that their poloidal field should consist of a dipole and a octupole component accompanied by a very weak toroidal quadrupole. For initial conditions that have a higher multipole $\ell$ structure the attractor consists mainly of $\ell$ and $\ell+2$ poloidal components.    
\end{abstract}

\date{\today}
\maketitle

{\it Introduction.}--- The evolution of the magnetic field inside an electrically neutral conducting medium, where only one species of particles is available to carry the electric current, is described by the Hall drift. The Hall effect has attracted attention in astrophysics as it can drive magnetic field evolution in neutron star (NS) crusts. This is because the crust is a highly conducting ion crystal lattice, where free electrons carry the electric current while any Lorentz force is balanced by elastic forces. Thus the system is always in dynamical equilibrium and its behaviour is described kinematically. 

A puzzling question in this context is whether the Hall effect leads to turbulent cascade \citep{Goldreich:1992} and complete dissipation of the field or whether there is a stable attractor state towards which the field relaxes. 2-D and 3-D simulations in cartesian boxes \citep{Wareing:2009, Cho:2009, Biskamp:1996}, support that the field undergoes turbulent cascade, although with evidence of stationary structures \citep{Wareing:2009b, Wareing:2010}. Turbulence develops even in the case of relatively low magnetic Reynolds numbers, $R_{M}\sim 30$. Supporting this, \cite{Lyutikov:2013} argued that any stationary closed configuration is neutrally stable and therefore would not be an attractor. This is unlike MHD where it is possible for the field to exchange energy with the plasma and evolve to a lower energy state. Crust studies assuming axial symmetry, on the contrary, do not find any sign of turbulent cascade, yet Hall evolution is non-trivial \citep{Shalybkov:1997, Hollerbach:2002, Hollerbach:2004, Pons:2009, Vigano:2012, Gourgouliatos:2014}. After some initial oscillatory behaviour which lasts longer for larger initial $R_{M}$ the Hall effect saturates \citep{Pons:2007}. This saturation occurs while the ohmic timescale is still much longer than the Hall, meaning that the evolution is still far from pure ohmic decay.

In this work we show that there is indeed an attractor state for axially symmetric magnetic fields evolving under the Hall effect and ohmic decay. This structure is characterised by constant electron angular velocity $\Omega$ along poloidal field lines, labeled by poloidal magnetic flux $\Psi$, similar to Ferraro's law \citep{Ferraro:1937}, with the additional property that $\Omega\approx \alpha \Psi$, where $\alpha$ is a constant. We find that a great variety of initial conditions of magnetic fields relax to this state, which on a longer timescale decays ohmically, retaining its structure.

{\it Hall Evolution.}--- An axially symmetric magnetic field in spherical coordinates $(r, \theta, \phi)$ can be written as $\bm{B}=\nabla \Psi \times \nabla \phi + I \nabla \phi$ where $I$ is related to the toroidal field and $cI(r, \theta)/2$ is the poloidal current passing through a spherical cap of radius $r$ and opening angle $\theta$. This field resides inside a conductor where only electrons of number density $n_{\rm e}$ are free to move with velocity $\bm{v}$. The electric current density is $\bm{j}=-{\rm e} n_{\rm e} \bm{v}$, the electric field is $\bm{E}= - \frac{\bm{v} \times \bm{B}}{c} +\frac{\bm{j}}{\sigma}$, where $\sigma$ is the electric conductivity, $c$ and ${\rm e}$ are the speed of light and the elementary electron charge. Using Amp\`ere's law $\nabla \times \bm{B} =(4\pi/c) \bm{j}$ the induction equation becomes: 
\begin{eqnarray}
\frac{\partial \bm{B}}{\partial t} = -\frac{c}{4 \pi{\rm e}}\nabla \times  \left(\frac{\nabla \times \bm{B}}{n_{\rm e}} \times \bm{B}\right) -\frac{c^{2}}{4\pi} \nabla \times \left(\frac{\nabla \times \bm{B}}{\sigma}\right)\,.
\label{dB}
\end{eqnarray}
The first term in the right-hand-side describes Hall evolution, the second ohmic dissipation. Their ratio is measured by the magnetic Reynolds number $R_{M}=\sigma |B|/(n_{\rm e} {\rm e}c)$. A magnetic field is in Hall equilibrium when the Hall term is zero \citep{Cumming:2004, Reisenegger:2007, Gourgouliatos:2013a, Gourgouliatos:2013b}. To understand the evolution of the magnetic field it is more illuminating to write equation \eqref{dB} in terms of $\Psi$ and $I$. To do so we define  $\chi=c/(4\pi {\rm e}n_{\rm e} r^{2}\sin^{2}\theta)$, the Grad-Shafranov operator $\Delta^{*}=\frac{\partial^{2}}{\partial r^{2}} +\frac{\sin\theta}{r^{2}}\frac{\partial}{\partial \theta}\left(\frac{1}{\sin\theta}\frac{\partial}{\partial \theta}\right)$ \citep{Reisenegger:2007}, and the electron angular velocity $\Omega=v_{\phi}/(r\sin\theta)= \chi \Delta^{*}\Psi$. Equation \eqref{dB} now reads 
\begin{eqnarray}
\frac{\partial \Psi}{\partial t} +r^{2}\sin^{2}\theta\chi (\nabla I \times \nabla \phi)\cdot \nabla \Psi =\frac{c^{2}}{4 \pi \sigma}\Delta^{*}\Psi\,,
\label{dPSI}
\end{eqnarray}
\begin{eqnarray}
\frac{\partial I}{\partial t} +r^{2}\sin^{2}\theta [\left( \nabla  \Omega \times \nabla \phi \right) \cdot \nabla \Psi 
+ I \left(\nabla \chi \times \nabla \phi \right) \cdot \nabla I ]  \nonumber \\
= \frac{c^{2}}{4 \pi \sigma}\left(\Delta^{*}I-\frac{1}{\sigma}\nabla I\cdot \nabla \sigma\right) \,.
\label{dI}
\end{eqnarray}  
The Hall evolution of the poloidal part of the field is mediated through the toroidal part, equation \eqref{dPSI}, while the toroidal part evolves either because of twisting of the poloidal field lines when $\Omega$ is not constant along a field line, or because of the geometric-density term $\chi$ if $I\neq I(\chi)$, equation \eqref{dI}. 

We used the code presented in \cite{Gourgouliatos:2014}. The initial maximum $R_{M}$ in the simulations was chosen to be in the range 40--80, and the evolution saturated after a few Hall timescales, with the maximum $R_{M}$ still being in the range of 20--50. The crust covers $0.2$ of the NS's radius, the electron density varies by two orders of magnitude from the surface of the star to the crust-core interface. Similar results have been found for a variety of other choices.

{\it The Hall Attractor}--- In \citep{Gourgouliatos:2014} we found the surprising result that for several different initial conditions the field adopted a similar state at late times, with Hall drift enforcing  isorotation, saturating the Hall effect. In this present work we show that this is a fundamental behaviour under the influence of Hall effect. After some initial Hall evolution the field relaxes to a particular isorotation profile characterized by $\Omega \approx \alpha \Psi$, where $\alpha$ is a constant of proportionality, which is insensitive to the choice of initial conditions. This result holds even if a higher multipole $\ell$ initial state is chosen, with the system relaxing to a mixture of $\ell$ and $\ell+2$, with $\Omega$ and $\Psi$ linearly related. Note that even if the system starts from some different isorotating profile it evolves to this particular one. 

\begin{figure}
\includegraphics[width=\columnwidth]{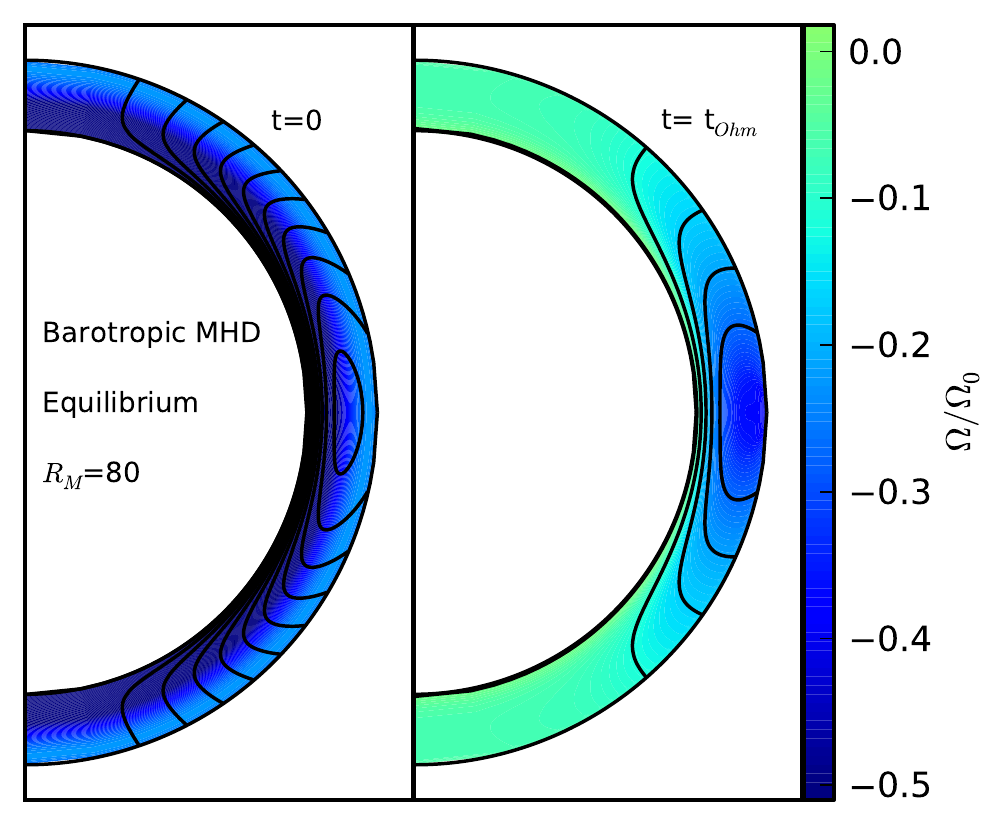}
\includegraphics[width=\columnwidth]{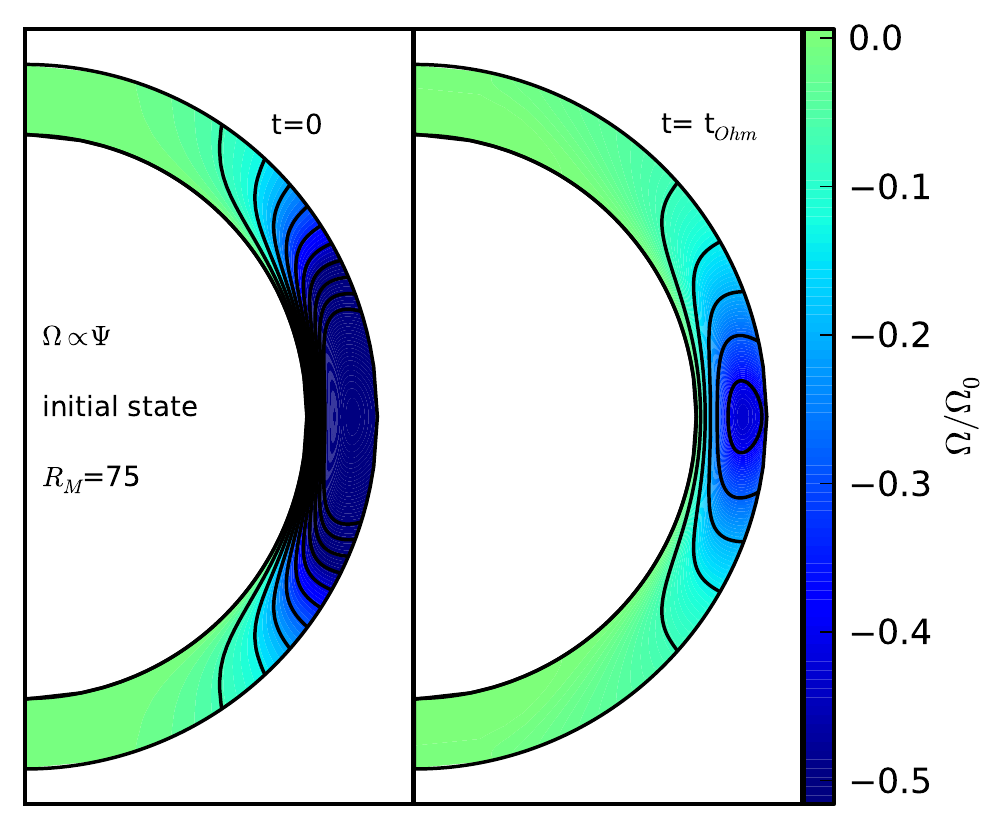}
\caption{A meridional section of the star showing $\Omega$ in color normalized to its maximum initial value $\Omega_{0}$ and the poloidal field lines in black. Top Panel: The initial state is a dipole poloidal field in barotropic MHD equilibrium with the electron fluid moving faster near the crust-core boundary. This field evolves towards isorotation as shown in the right plot where contours of constant $\Omega$ coincide with those of constant $\Psi$. Bottom Panel: The initial state is chosen to be $\Omega = \alpha \Psi$, with $\alpha$ a negative constant, the field starts in isorotation and maintains this state while dissipating. }
\label{Fig:1}
\end{figure}

\begin{figure}
\includegraphics[width=\columnwidth]{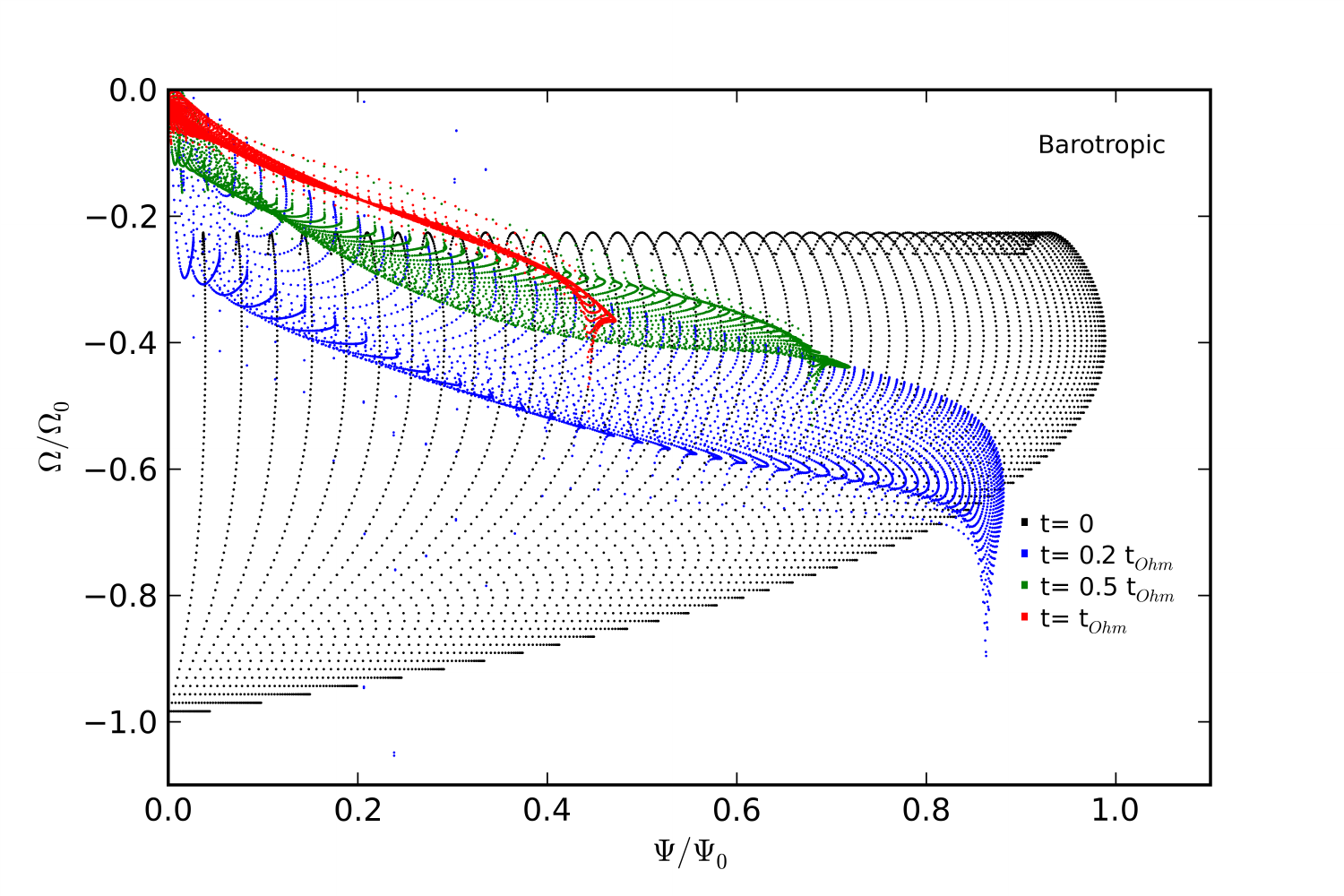}
\includegraphics[width=\columnwidth]{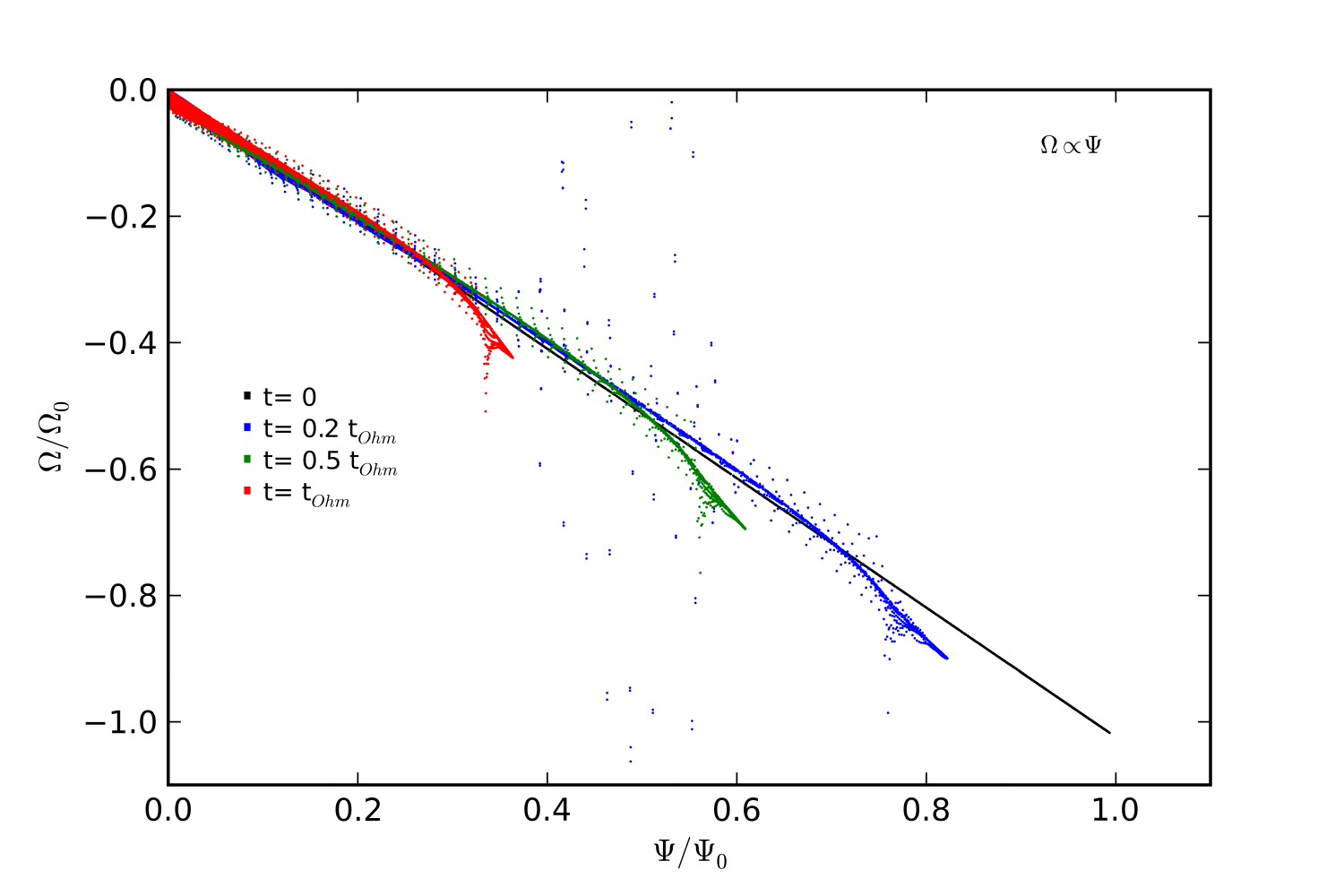}
\caption{Scatter plot of $\Omega$ and $\Psi$, each point corresponds to a grid point of the simulation. Top Panel: The scatter plot for the case shown in Figure \ref{Fig:1} top panel. Initially there is differential rotation of the electron fluid along the field lines (black points), with multiple values of $\Omega$ for given $\Psi$, which after some Hall evolution tend to concentrate in a narrower region (blue and green points), and eventually the system saturates to isorotation (red points). Bottom Panel: The scatter plot for a system starting with $\Omega=\alpha \Psi$ (black points), the initial structure is very close to the attractor state, thus the system changes only slightly its structure as shown with $\Omega$ and $\Psi$ deviating from linearity (blue and green points), however the system maintains isorotation, even after a significant part of the field has been dissipated (red points). }
\label{Fig:2}
\end{figure}

We explored a wide variety of initial conditions and crust profiles, including cases out of Hall equilibrium, mixed initial poloidal and toroidal fields of various polarities and energy ratios. In general, the early evolution is a response to any imbalanced terms, with whistler waves launched. Eventually the system relaxes to a state where $\Omega \approx \alpha \Psi$, the details of the relation depend on the choice of the density profile and conductivity. An example is shown Figure \ref{Fig:1} (top panel) which starts with significant differential rotation but is in a state of isorotation after one ohmic time. To make this evident we plotted ($\Omega, \Psi$) for every grid point of our simulation, Figure \ref{Fig:2}.

We noticed that the Hall attractor $\Omega(\Psi)$ lies close to the minimum rotation rate of the lowest order ohmic mode. This naturally gives $\Omega \propto \Psi$ in the attractor state since the angular velocity for an ohmic eigenmode $\Omega=\chi \Delta^{*}\Psi= 4 \pi\sigma \chi/(c^{2}\tau) \Psi$, where $\tau$ is the decay time.  Although the ohmic mode has significant differential rotation due to the variation of $\sigma \chi$ across the crust, we find that we can approximately reproduce the attractor by choosing the lowest value of $\sigma \chi$ along each field line. As most of the field lines cross the equator near the base of the crust they have the same minimum value of $\sigma \chi$, therefore the $\Omega$-$\Psi$ relation is almost linear.


To further investigate the properties of this state we constructed a Hall equilibrium where $\Omega \propto \Psi$ and $I=0$, and used it as an initial condition, for various choices of $n_{\rm e}$ and $\sigma$ profiles. Once Hall evolution starts, the dependance of $\Omega(\Psi)$ changes slightly while maintaining its isorotation and its overall structure, Figures \ref{Fig:1} and \ref{Fig:2} bottom panels. We decompose the magnetic field on the surface of the star in terms of $c_{\ell}=(2\ell +1)/(2\ell +2) \int_{-1}^{1} B_{r} P_{\ell} (\mu) d\mu$, $B_{r}$ is the radial field on the NS's surface, $\mu=\cos\theta$ and $P_{\ell}$ the Legendre Polynomial of $\ell$th order. The Hall attractor field consists of a dipole component ($\ell=1$), and an octupole ($\ell=3$) whose relative intensity depends on the crust properties, and is counter-aligned with the dipole. Higher multipoles are present, but their intensity is smaller. The weak toroidal field developed is in the $\ell=2$ component.  
\begin{figure}
\includegraphics[width=\columnwidth]{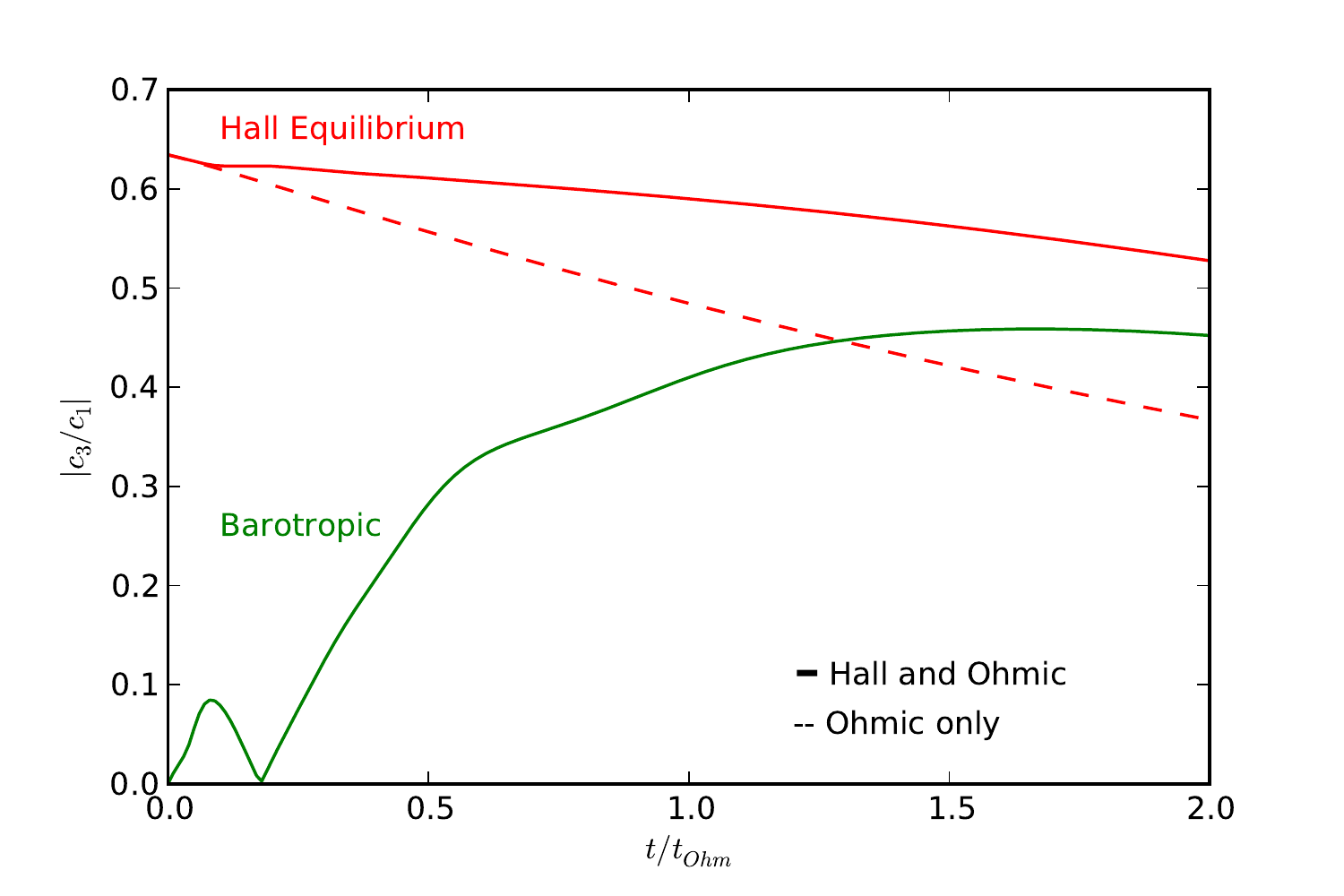}
\caption{The ratio of the octupole over the dipole component on the surface of the star, as a function of time, for a magnetic field that starts at $\Omega = \alpha \Psi$ and evolves under the Hall Effect or ohmically only red solid and dashed curves respectively, and for a barotropic equilibrium initial condition shown in green. Under the influence of ohmic dissipation only, the octupole decays faster than the dipole component, thus their ratio decreases, whereas when the Hall term is included, after some transient initial evolution the ratio stays almost constant. The field that started from the barotropic initial condition eventually reaches a ratio close to the $\Omega=\alpha \Psi$ state.}
\label{Fig:3}
\end{figure}

The finite conductivity leads to some dissipation of the field. As the ohmic dissipation timescale for the $\ell=3$ poloidal component is shorter than that of the  $\ell=1$, it pushes the magnetic configuration out of the isorotation state. The role of the $\ell=2$ toroidal field is to transfer energy from the $\ell=1$ poloidal component into the $\ell=3$ to compensate for the losses. Indeed the ratio of the $\ell=1$ and $\ell=3$ varies slowly, as opposed to a pure ohmic decay, where eventually $\ell=1$ dominates, Figure \ref{Fig:3}. This evolution is an advection-diffusion equilibrium where the system maintains its structure and evolves self-similarly with time, with the energy being dissipated by the ohmic term and the slightly imbalanced Hall term rearranging the field so that the changes in the structure are annulled.  

\begin{figure}
\includegraphics[width=\columnwidth]{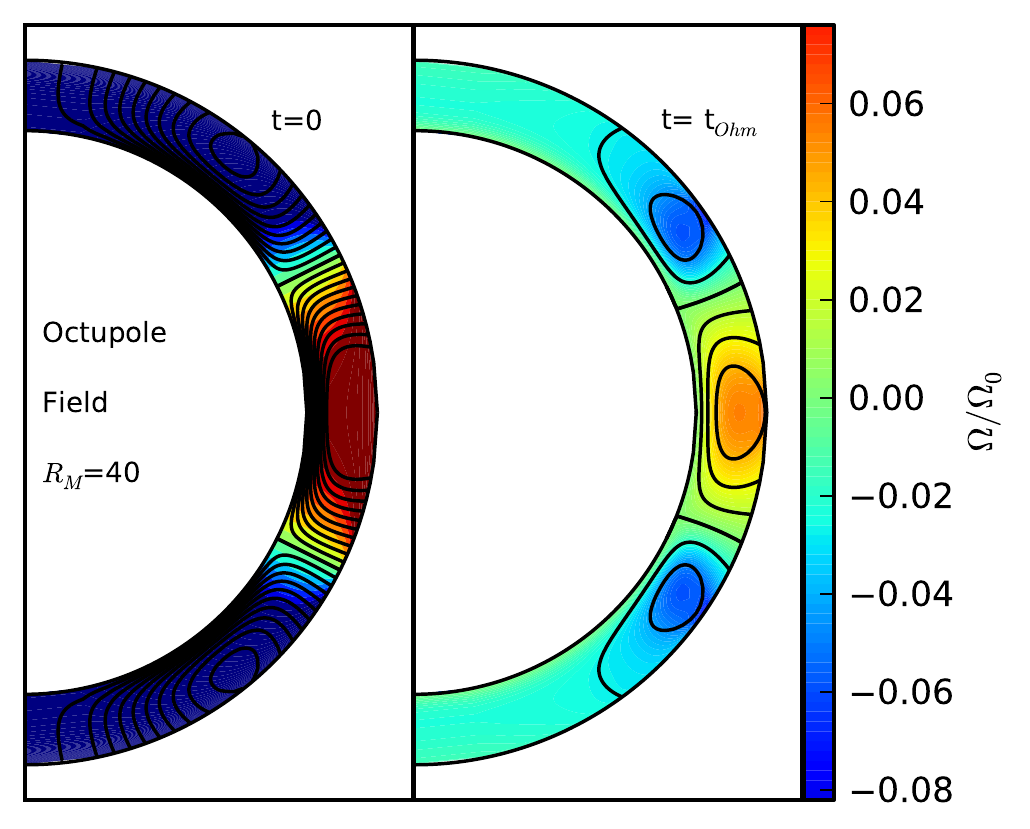}
\includegraphics[width=\columnwidth]{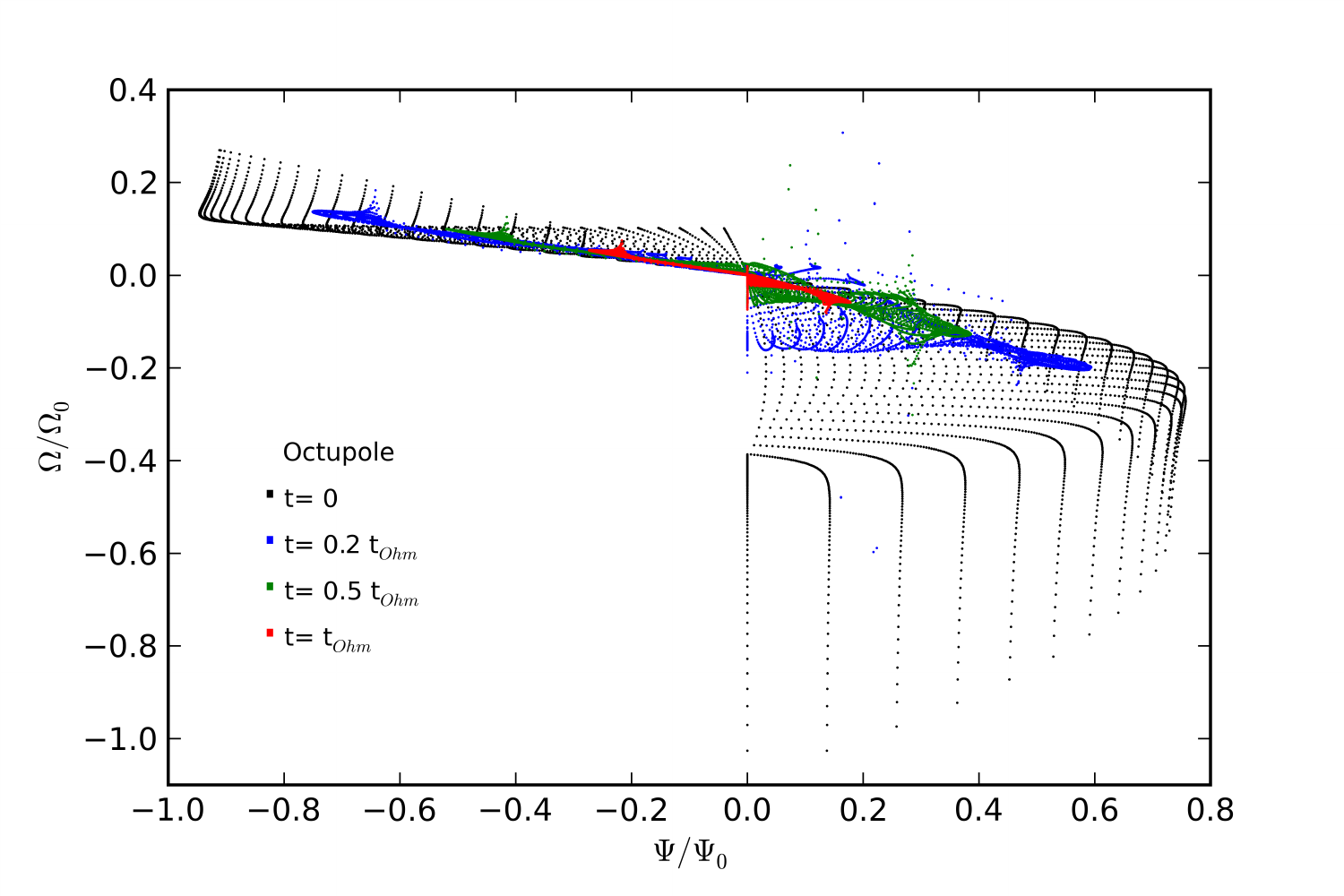}
\caption{The structure of an octupole field  and $\Psi, \Omega$ scatter. The field reaches isorotation, eventually, even if the initial state is dominated by a higher $\ell$.}
\label{Fig:4}
\end{figure}

Quite remarkably, a choice of an initial poloidal field consisting of $\ell=2$, $\ell=3$ or a higher multipole leads to a long-term state which is dominated by the initial component and an  $\ell+2$ multipole for the poloidal field while the toroidal field is of multipole order $\ell+1$. The field tends to relax to an isorotation state with approximate linear relation between $\Psi$ and $\Omega$, as shown for an $\ell=3$ field in Figure \ref{Fig:4}.

Even in the case of constant density, the dependence of $\chi$ on $r^{2}\sin^{2}\theta$ differentiates the magnetic evolution of a cartesian box simulation to a crust. To test that hypothesis, we simulated crusts with $n_{\rm e} \propto 1/(r^{2}\sin^{2}\theta)$, thus $\chi=$const.~and compared them with realistic $\chi\neq$const.~simulations. We found that higher multipoles were developing much faster and to higher intensities in $\chi=$const.~simulations, compared to the $\chi\neq$const.~ones. This difference is mainly because of the second term of equation \eqref{dI}, indeed if we expand it we find $r^{2}\sin^{2}\theta[(\chi \nabla \Delta^{*}\Psi + \Delta^{*}\Psi \nabla \chi) \times \nabla \phi]\cdot \nabla \Psi$. Running multiple simulations we noticed that these two terms tend to have opposite contributions. On the contrary, the third term in the right-hand-side of equation \eqref{dI} which was quadratic in $I$ is in general weak and has a minor contribution. This can provide a path to compare the behaviour of the magnetic field in cartesian box simulations and crusts, however further investigation is required.

{\it Neutron Star Long Term State}---  A newborn NS undergoes a stage during which its crust freezes, initiating Hall evolution. A realistic crust of thickness $1$km, $n_{\rm e}$ ranging from $10^{32-36}$cm$^{-3}$, $\sigma$ in the range $10^{21-24}$s$^{-1}$ and an initial surface magnetic field of $10^{14}$G, the NS needs $\sim 1$ Myr to evolve towards the attractor state. Once it reaches the attractor state $R_{M}$ is still significantly larger than unity and spends a few Myr of its life in this state, until the field has dissipated so much that the ohmic timescale is comparable with the Hall. Once the Hall and ohmic timescales are comparable higher multipoles dissipate faster with the dipole one surviving the longest. 

This has important implications for the field structure of middle aged NSs whose magnetic field exceeds $5\times 10^{12}$ G. Their surface magnetic field should consist mainly of a dipole and an octupole with a ratio of octupole to dipole about $\sim$2/3 and opposite polarity, severely altering the idealized picture of the dipole field. Given that the spin-down calculation takes into account only the dipole component of the field, the intensity of the magnetic field at the polar cap including the higher multipoles should be $\sim$1/3 of the intensity of what the dipole model predicts, while the equatorial field should be $\sim$2.5 times stronger, constraining the assumed models of the magnetic field. Fits of thermal profiles of isolated neutron stars have suggested an offset dipole or multipolar structure \cite{Haberl:2007}. Weaker magnetic field neutron stars with $B\lesssim 10^{12}$ G could also undergo a Hall-dominated phase of evolution if the crust has an impurity parameter significantly less than unity (see~Fig. 4 of \cite{Cumming:2004}), although with a much longer Hall timescale.

This result disfavours the idea that Hall effect leads to turbulent cascade of the magnetic field in neutron star crusts, as there is indeed an attractor state towards which the field is trying to relax, which requires the excitement of a higher order multipole. Additionally this is an example of a kinematic physics problem which has an attractor state, despite the fact that its evolution equation does not arise from an energy minimization principle. Given that these results were found in a system where axially symmetry is assumed, we stress the importance of the development of 3-D crust studies, either analytically or numerically to investigate whether the attractor persists in 3-D. In 3D, non-axisymmetric modes are available to participate in a cascade \citep{Lyutikov:2013}; on the other hand, in rotating stars it is possible to find non-axisymmetric generalizations of Ferraro's law involving isorotation with additional motion along field lines \citep{Mestel:1988, Spruit:1999, Charbonneau:1993}.

{\it Acknowledgments}--- We thank Dave Tsang, Vicky Kaspi, Toby Wood and Ulrich Geppert for insightful discussions. KNG was supported by the Centre de Recherche en Astrophysique du Qu{\'e}bec.  AC is supported by an NSERC Discovery Grant and is an Associate Member of the CIFAR Cosmology and Gravity program.

Animations of the evolution of the magnetic field lines, the electron velocity and $(\Psi,\Omega)$ can be found at: \url{http://www.physics.mcgill.ca/~kostasg/research.html}.

\bibliography{BibTex.bib}

\end{document}